\documentclass[runningheads]{llncs}
\usepackage[T1]{fontenc}
\usepackage{graphicx}
\usepackage{amsmath} 
\usepackage{bbm}
\usepackage{cite}
\usepackage{soul}
\usepackage{color}
\usepackage{ulem}
\usepackage{multirow}
\usepackage{graphicx}
\usepackage{marvosym}
\usepackage{cleveref}
\begin{document}
\title{Hybrid Kinetics Embedding Framework \\ for Dynamic PET Reconstruction}
\titlerunning{Hybrid Kinetics Embedding Framework}
%
%
% If the paper title is too long for the running head, you can set
% an abbreviated paper title here
%
% \author{Anonymous Authors}
\author{
Yubo Ye\inst{1}, 
Huafeng Liu\inst{1,3,4} \textsuperscript{(\Letter)}, 
and Linwei Wang\inst{2}
}
%\author{First Author\inst{1}\orcidID{0000-1111-2222-3333} \and
% Second Author\inst{2,3}\orcidID{1111-2222-3333-4444} \and
% Third Author\inst{3}\orcidID{2222--3333-4444-5555}}
%
% \authorrunning{Anonymous Authors}
\authorrunning{Y. Ye et al.}

\institute{
State Key Laboratory of Modern Optical Instrumentation, \\
Department of Optical Engineering, Zhejiang University, Hangzhou 310027, China \\
\email{liuhf@zju.edu.cn} \and
Rochester Institute of Technology, Rochester, NY 14623, USA \and
Jiaxing Key Laboratory of Photonic Sensing \& Intelligent Imaging, \\
Jiaxing 314000, China \and
Intelligent Optics \& Photonics Research Center, Jiaxing Research Institute, \\
Zhejiang University, Jiaxing 314000, China
}
% First names are abbreviated in the running head.
% If there are more than two authors, 'et al.' is used.
%
%\institute{Princeton University, Princeton NJ 08544, USA \and
%Springer Heidelberg, Tiergartenstr. 17, 69121 Heidelberg, Germany
%\email{lncs@springer.com}\\
%\url{http://www.springer.com/gp/computer-science/lncs} \and
%ABC Institute, Rupert-Karls-University Heidelberg, Heidelberg, Germany\\
%\email{\{abc,lncs\}@uni-heidelberg.de}}
%
\maketitle  % typeset the header of the contribution
\begin{abstract}
%Dynamic positron emission tomography (PET) imaging can provide 
%spatial-temporal information about biochemical and physiological processes in \textit{vivo}, used for kinetic analysis and auxiliary diagnosis. 
In dynamic positron emission tomography (PET) reconstruction, 
%Compared to static PET reconstruction, 
the importance of leveraging the temporal dependence of the data has been well appreciated. 
Current deep-learning solutions can be categorized in two groups in the way the temporal dynamics is modeled: 
data-driven approaches use spatiotemporal neural networks to learn the temporal dynamics of tracer kinetics from data, which relies heavily on data supervision; 
physics-based approaches leverage \textit{a priori} tracer kinetic models to focus on inferring their parameters, 
%to estimate the parameters parametric image of prior tracer kinetic model and then achieve reconstruction, 
which relies heavily on the  
accuracy of the prior kinetic model. 
In this paper, we marry the strengths of these two approaches in a hybrid kinetics embedding  (HyKE-Net) framework for dynamic PET reconstruction. We first introduce a novel \textit{hybrid} model of tracer kinetics consisting of 
a physics-based function augmented by a neural component to account for its gap to data-generating tracer kinetics, both identifiable from data. 
We then 
%physics-based kinetic model as a universal differential equation, make it conditioned on additional neural kinetic parameter and then 
embed this hybrid model at the latent space of an encoding-decoding framework 
to enable both supervised and unsupervised 
identification of the hybrid kinetics and thereby dynamic PET reconstruction. 
%into data-driven reconstruction frame. 
Through both phantom and real-data experiments, we demonstrate the benefits of HyKE-Net -- especially in unsupervised reconstructions -- over
existing physics-based and data-driven baselines as well as its ablated formulations where the embedded tracer kinetics are purely physics-based, purely neural, or hybrid but with a non-adaptable neural component. 
%embedding hybrid kinetics, which integrates rich prior knowledge while identifying its gap to observed data.

\end{abstract}

\section{Introduction}
Dynamic positron emission tomography (PET) imaging 
is widely used in clinical diagnosis and treatment 
for elucidating the spatial distribution and kinetics of radiotracer-labeled biological substrates \textit{in-vivo} \cite{apply1, apply2, apply3}. %\textcolor{blue}{
Dynamics PET reconstruction aims to reconstruct a temporal series of spatial concentration distribution of the  radioactive tracer (tracer activity images) from a series of noisy sinograms measured by a PET scanner. %}
%\hl{Needs a sentence to define dynamic PET reconstruction, introducing key terms including the input/measurement data and the output of interest.}
This reconstruction remains a challenging task due to the complex spatio-temporal nature of the tracer kinetics as well as the high level of noise and low counting statistics in the sinogram measurements \cite{difficulty1,difficulty2}, all contributing to the ill-posedness of this inverse problem \cite{ill-posedness}.

Compared to static PET reconstruction \cite{FBP, Iter1, Iter2, Pena1, Pena2, Pena3}, the importance to leverage the temporal dynamics of the data is well appreciated in dynamic PET reconstruction. % \cite{dPET1, dPET2, dPET3}. 
%\textcolor{blue}{
While various traditional optimization and inference strategies have been introduced to incorporate temporal prior as constrains for dynamic PET reconstruction \cite{dPET1, dPET2, dPET3, kinetic3}, recent deep learning solutions have shown state-of-the-art performances % for this task. %}
%\hl{While various traditional optimization and inference strategies have been introduced to leverage tracer kinetics as prior constrains to dynamic PET reconstruction} \cite{dPET1, dPET2, dPET3}, recent
% deep learning methods \hl{have shown state-of-the-art performances for this task?}
%show extraordinary capabilities in medical imaging and kinetic modeling, many deep learning-based dynamic PET reconstruction methods have been proposed, 
%Existing related works 
and can be categorized into 
data-driven \textit{vs}.\ physics-based approaches to modeling the tracer kinetics underlying the data. 

%learning of the governing 
%kinetics, and physics-based estimation of the parameters of \textit{a priori} tracer kinetic models.
% \textcolor{blue}{voxel-wise parameters of 
% \textit{a priori} tracer kinetic model (parametric image)} \hl{define parametric image}.

The data-driven approaches \cite{learning1, learning2, learning3, learning4} generally leverage spatio-temporal neural networks to learn kinetic information directly from data.
%\textcolor{blue}{
The learning of this \textit{black-box} kinetics, 
however, neglects potentially valuable prior knowledge about the tracer kinetics and relies heavily on the supervision of the ground truth tracer activity images 
-- the latter 
are difficult to obtain in practice. %} 
%\hl{[what? this should be defined in the sentence added in paragraph 1] [need to elaborate a bit on how these supervising data are obtained; if using simulation data, what is the challenge in real data use, etc.] }
%This neglects potentially valuable prior knowledge about the tracer kinetics, 
%\hl{while limiting their applicability in \textit{in-vivo} settings where [something related to the data requirement tied to the added elaboration above.]}
Alternatively, physics-based approaches \cite{kinetic1, kinetic2} 
leverage \textit{a priori} tracer kinetic model in known mathematical expressions and 
focus on estimating its pixel-wise
% voxel-wise 
parameters (parametric images)
%parametric images of the model 
in order to reconstruct tracer activity images.
%\hl{[I don't know what is parameteric images..again related to the definition in paragraph.]}\textcolor{red}{Every pixel in a image has its kinetic parameters and all these paramters make up a parameteric image} \hl{is activity image different from parametric image? If different, parametric image needs to be defined the first time it is used.} 
This \textit{white-box} kinetics 
introduces valuable prior physics to reduce the reliance on data, 
although the performance of these approaches can decline as the gap between the prior and data-generating kinetics increases.  
%the performance of these methods will decline. 
How to effectively leverage physics-based \textit{a priori} kinetic models 
to remove the reliance on ground-truth tracer activity images, 
while addressing its unknown gap to data-generating tracer kinetics remains an open question in dynamic PET reconstruction. 

In this paper, we answer this important question with a novel \textit{hybrid} kinetics-embedding framework (HyKE-Net) for dynamic PET reconstruction.
%\hl{(Hyke?)} 
As outlined in Fig.~\ref{fig:overview}, 
HyKE-Net has two important innovations. 
First, we model the tracer kinetics with a novel hybrid expression, in the form of a universal differential equation \cite{UDE} that consists of both a known mathematical expression of tracer kinetics representing prior knowledge and a neural function representing its gap to actual data-generating tracer kinetics.  
Second, we embed this hybrid tracer kinetics in the latent space of an encoding-decoding architecture, enabling (through the encoder) separate identification of both its prior physics-based and unknown neural components directly from measured sinograms, 
while allowing (through the decoder) \textit{unsupervised} training with only a reconstruction objective at the sinogram space without access to the ground-truth tracer activity images.  

We evaluated HyKE-Net for both unsupervised and supervised dynamic PET reconstruction on synthetic and real datasets, in comparison to two traditional methods (FBP \cite{FBP} and Joint-TV \cite{kinetic3}) and one data-driven deep-learning method (FBP-Net \cite{learning2}). 
We further ablated the use of hybrid kinetics with the use of purely physics-based, purely neural, and hybrid kinetics but with a global neural component in HyKE-Net. The results provide strong evidence for the benefits of embedding hybrid tracer kinetics to integrate rich prior knowledge while allowing for its errors, especially in unsupervised dynamics PET reconstruction.

% \hl{What conclusions on both unsupervised tasks, and supervised tasks.}

\section{Background}
\subsubsection{PET Imaging Model:}
\label{Imaging Model}
The relationship between tracer activity images $\mathbf{x}$ and measured sinograms $\mathbf{y}^r$ can be described by:
\begin{equation}
    \label{eq:projection}
   \mathbf{y}^r \sim Poisson( \Bar{\mathbf{y}}= 
   \mathbf{D}(\mathbf{G}\mathbf{x})+\mathbf{n})
\end{equation}
where $\mathbf{G}$ describes the 
discrete radon transform (\textit{i.e.}, projection),
% followed by the measurement process by the PET scanner:
% where $\mathbf{G}$ denotes the projection matrix,
%$\mathbf{y}^r$the raw sinograms measured by the PET scanner, 
%are further degraded by the following measurement model:
% \begin{equation}
%     \label{eq:measurement}
%     \mathbf{y}^r \sim Poisson(\Bar{\mathbf{y}}= \mathbf{D}\mathbf{y}+\mathbf{n})
% \end{equation}
%where 
%measurement matrix 
$\mathbf{D}$ models %\textcolor{blue}{
the detection-probability of the PET scanner, %}
%effects such as photon attenuation, detector efficiency, and detection-probability, 
noise matrix $\mathbf{n}$ represents the expectation of randoms and scatters, 
and $\Bar{\mathbf{y}}$ the expectation of measured sinograms.
% random photons and scatter photons. 
%\textcolor{blue}{This description is accurate in PET, I think} 
% discreted \hl{discrete? although not sure what is discreate in the equation below} 
% \begin{equation}
% \label{eq:projection}
%    \text{Projection: } p(r,\varphi)=\int\limits_{-\infty}^{\infty}\int\limits_{-\infty}^{\infty}f(x,y)\delta(x\cos\varphi+y\sin\varphi-r)dxdy
% \end{equation}
%The raw data 
% $\mathbf{y}^r$
%measured by PET scans \hl{[is "measured by PET scans" a proper phrase?]} are then degraded sinograms which can be expressed 

\subsubsection{Tracer Kinetics of Dynamic PET:}
Various models have been proposed to model tracer kinetics in dynamic PET. Due to the simple implementation and biological plausibility, 
%in this paper, 
we consider the one- and two-tissue compartment models:
\begin{align}
\label{eqn:1cmpt}
    \text{One-tissue: }& \dot{C}_{Ti}(t)=-k_{2i}C_{Ti}(t)+k_{1i}C_P(t) \\ \nonumber 
    \text{Two-tissue: }& \begin{bmatrix}\dot{C}_{Ei}(t) \\ \dot{C}_{Mi}(t) \end{bmatrix}=\begin{bmatrix}-k_{2i}-k_{3i} & k_{4i}\\ k_{3i} & -k_{4i} \end{bmatrix}
                  \begin{bmatrix}C_{Ei}(t) \\ 
                     \label{eqn:2cmpt} C_{Mi}(t)\end{bmatrix}+
                  \begin{bmatrix}k_{1i} \\ 0 \end{bmatrix}C_P(t) \\
                  & C_{Ti}(t)= {C}_{Ei}(t)+{C}_{Mi}(t)
\end{align}
where $C_P$ is the arterial concentration of nonmetabolized tracer in plasma, $C_E$ is the concentration of nonmetabolized tracer in tissue, $C_M$ is the concentration of the radioisotope-labeled metabolic products in tissue, $C_T$ is the total concentration of
radioactivity in tissue, subscript $i$ denotes different pixel
% voxel 
locations, and $k_1,k_2,k_3,k_4$ are physical kinetic parameters. 
The estimated values of these parameters across pixels
%voxels 
provide the \textit{parametric images} in PET reconstruction. 
The \textit{tracer activity image} $\mathbf{x}_k$ at the $k$th time frame can then be computed as:
\begin{equation}
\label{eqn:forwardimage}
\mathbf{x}_k =
\left \{ \frac{1}{t_k-t_{k-1}} \int_{t_{k-1}}^{t_k}
\left [ (1-V_{i})C_{Ti}(t)+V_{i}C_P(t)  \right ] e^{-\frac{t}{\tau}}dt \right \}_{i=1:W\times H}
\end{equation}
where $\tau$ is the tracer decay coefficient, $t_k-t_{k-1}$ the scan interval, $V$ the vascular volume fraction, and 
%$N=W\times H$ where 
$W$ and $H$ the width and height of the %2D 
tracer activity image.
%\hl{so you refer to i as a voxel implying that this is a 3D stack of images, but here N is 2D?}

\section{Methodology}
\begin{figure}[t]
    \centering
    \includegraphics[width=\linewidth]{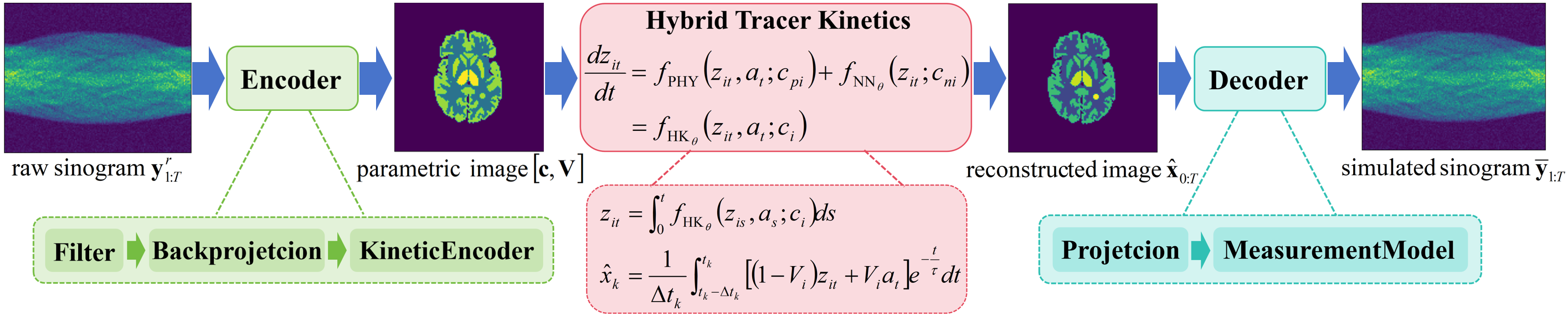}
    \caption{Overview of HyKE-Net with hybrid tracer kinetics in the latent space to facilitate supervised or unsupervised dyanmic PET reconstruction.}
    \label{fig:overview}
\end{figure}

As illustrated in Fig.~\ref{fig:overview}, 
HyKE-Net has two key design components: 
1) an embedded hybrid tracer kinetic model combining a prior physics-based compartment model and an unknown neural function describing errors in prior physics, and 
2) an encoding-decoding formulation that bridges this embedding space to the space of tracer activity images and further to the space of measured sinograms, to enable unsupervised identification of hybrid tracer kinetics and thereby unsupervised reconstruction of 
the time sequence of tracer activity images.
%dynamic PET reconstruction.

%The proposed hybrid kinetics embedding framework consists of three parts: FBP part, kinetics part and imaging part. 

\subsubsection{Hybrid Tracer Kinetics:} 
% \hl{you should describe just the hybrid kinetics function here; no inference or anything}
We first describe the tracer kinetics as a hybrid model in the form of a universal differential equation \cite{UDE}, with a 
physics-based compartment model augmented by a neural kinetic function:
\begin{align}
    \frac{d\mathbf{z}_{it}}{dt} &=
%    -k_{2i}z_{it}+k_{1i}a_t+f_{\text{NN}_\theta}(z_{it};c_{ni}) \\
    f_{\text{PHY}}(\mathbf{z}_{it}, a_t; \mathbf{c}_{pi}) + f_{\text{NN}_\theta}(\mathbf{z}_{it};\mathbf{c}_{ni}) 
     =f_{\text{HK}_\theta}(\mathbf{z}_{it},a_t;\mathbf{c}_i)\\ 
%     \end{align} 
%and its generation to tracer activity image as: 
%\begin{align}
\label{eqn:odesolve}
    \mathbf{z}_{it} & =\int_{0}^{t}f_{\text{HK}_\theta}(\mathbf{z}_{is},a_s;\mathbf{c}_i)ds% \\
%     \label{eqn:forwardimage}
%     \mathbf{x}_{1:T} &=
%     \left \{ \frac{1}{\Delta t_{k}} \int_{t_{k}-\Delta t_{k}}^{t_{k}}
% \left [ (1-V_{i})
% \mathbf{z}_{it}  +V_{i}a_t  \right ] e^{-\frac{t}{\tau}}dt \right \}_{i=1:N,k=1:T}
\end{align}
% \begin{equation}
%     \mathbf{z}_{it}=\int_{0}^{t}f_{\text{HK}_\theta}(\mathbf{z}_{is},a_s;\mathbf{c}_i)ds
% \end{equation}
% \begin{equation}
%     \hat{\mathbf{x}}_{1:T}=
%     \left \{ \frac{1}{\Delta t_{k}} \int_{t_{k}-\Delta t_{k}}^{t_{k}}
% \left [ (1-V_{i})\mathbf{z}_{it}  +V_{i}a_t  \right ] e^{-\frac{t}{\tau}}dt \right \}_{i=1:N,k=1:T}
% \end{equation}
where 
$f_{\text{PHY}}(\mathbf{z}_{it}, a_t; \mathbf{c}_{pi})$ represents either the one- or two-tissue compartment model introduced in Equations \eqref{eqn:1cmpt}-\eqref{eqn:2cmpt}, 
%\hl{is your c always ct?} \textcolor{blue}{Yes}
%$z_{it}=C_{Ti}(t)$, 
$a(t)=C_P(t)$, and $\mathbf{c}_{pi}$ include its physical kinetic parameters.  
$f_{\text{NN}_\theta}(\mathbf{z}_{it};\mathbf{c}_{ni})$ is a neural network parameterized by $\theta$, intended to account for the gap between the prior and data-generating tracer kinetics: 
%$c_{pi}=[k_{1i}, k_{2i}]$, $c_{ni}$ are physical and neural kinetic parameter respectively, $c_i=[c_{pi}, c_{ni}]$, $z_{it}=C_{Ti}(t)$ and $a(t)=C_P(t)$. 
%The neural kinetic function $f_{\text{NN}_\theta}(z_{it}; c_{ni})$ here is a simple fully-connected network (FCN) and 
instead of assuming a single neural function $f_{\text{NN}_\theta}(\mathbf{z}_{it})$ globally shared by all training sequences, 
we assume that it can vary with the training samples and set it 
conditioned on an additional abstract neural kinetic parameter $\mathbf{c}_{ni}$. $\mathbf{c}_{pi}$ and $\mathbf{c}_{ni}$
($[\mathbf{c}_{pi}, \mathbf{c}_{ni}]=\mathbf{c}_i$) will be inferred from individual sequences to allow the identification of hybrid tracer kinetics. 
The generation of tracer activity images $\mathbf{x}_{1:T}$ from $\mathbf{z}_{it}$ is as described in Equation \eqref{eqn:forwardimage}, in which $V_i$ will also be inferred. % to reconstruct tracer activity image underlying observed sinograms.
%Then, with well-identified $f_{\text{HK}_\theta}(\mathbf{z}_{it},a_t;\mathbf{c}_i)$, the reconstructed images $\hat{\mathbf{x}}_{0:T}$ can be computed as:

\subsubsection{Identification of Hybrid Tracer Kinetics:} 
% \hl{You should talk about both the FBP and context encoder here. }
%Instead of directly inferring kinetic parameter \hl{$[c_{ni}, V_i]$} from raw sinograms $\mathbf{y}_{1:T}^r$ in sinogram domain, 
% First, we construct a physics-embedded encoder to obtain rough images $\mathbf{\tilde{x}}_{1:T}$ in image domain from raw sinograms $\mathbf{y}_{1:T}^r$ in sinogram domain. 
To infer kinetic parameters  $[\mathbf{c}_i, V_i]$ from measured sinograms, instead of a purely neural encoder, we consider a hybrid encoder that transforms from the sinogram domain to image domain by a physics-embedded process. 
Inspired by the FBP algorithm \cite{FBP}, this is achieved via a filtering process with a learnable spatial filter $h_\phi(\cdot)$ parameterized by $\phi$ and a backprojection process with a backprojection matrix $\mathbf{G}^T$:
\begin{equation}
\label{eqn:BPF}
    \mathbf{\tilde{x}}_{1:T}=\mathbf{G}^T(h_\phi(\mathbf{y}_{1:T}^r))
\end{equation} 
\iffalse
\begin{equation}
    \text{Backprojection: } f(x,y)= \int\limits_{-\infty}^{\infty}\int\limits_{0}^{\pi}\hat{p}(r,\varphi)\delta(x\cos\varphi+y\sin\varphi-r)drd\varphi
\end{equation} 
\begin{equation}
\label{eqn:BPF}    
\mathbf{\tilde{x}}_{1:T}=\text{Backprojection}(h_\phi(\mathbf{y}_{1:T}^r))
\end{equation}
\fi

We then utilize a kinetic encoder $g_\varphi(\cdot)$ with paremeters $\varphi$ to infer kinetic parameter of each pixel (parametric images) as: 
%$\left[\mathbf{c}=\left \{ c_{i} \right \}_{1:N}, \textbf{V}=\left \{ V_{i} \right \}_{1:N} \right ]$ from $\mathbf{\tilde{x}}_{1:T}$.
\begin{equation}
    [\hat{\mathbf{c}}, \hat{\textbf{V}}] = g_\varphi(\mathbf{\tilde{x}}_{1:T}), \quad \hat{\mathbf{c}}=\left \{ \hat{\mathbf{c}}_{i} \right \}_{1:N}, \quad 
    \hat{\textbf{V}}=\left \{ \hat{V}_{i} \right \}_{1:N} 
\end{equation}
% \hl{it's better if you could spell these models out with some math symbols with its learnable parameters, so you can use these parameters in the final loss.}
% \hl{Also how about cp and cn -- are they coming from the same encoders?}

\iffalse
With inferred kinetic parameters 
$\hat{\mathbf{c}}_i=[\hat{\mathbf{c}}_{pi}, \hat{\mathbf{c}}_{ni}]$, 
\hl{or
$[c_{ni}, V_i]$? is v inferred as well? and what happens to cpi?}
the reconstructed images $\hat{\mathbf{x}}_{0:T}$ can be computed as:
\begin{equation}
    \hat{\mathbf{z}}_{it}=\int_{0}^{t}f_{\text{HK}_\theta}(\hat{\mathbf{z}}_{is},a_s;\hat{\mathbf{c}}_i)ds
\end{equation}
\begin{equation}
    \hat{\mathbf{x}}_{1:T}=
    \left \{ \frac{1}{\Delta t_{k}} \int_{t_{k}-\Delta t_{k}}^{t_{k}}
\left [ (1-V_{i})\hat{\mathbf{z}}_{it}  +V_{i}a_t  \right ] e^{-\frac{t}{\tau}}dt \right \}_{i=1:N,k=1:T}
\end{equation}
\fi

\subsubsection{Supervised and Unsupervised Reconstructions:}
Given inferred parameters $[\hat{\mathbf{c}}, \hat{\textbf{V}}]$, the tracer activity images $\hat{\mathbf{x}}_{1:T}$ can be reconstructed by Equations \eqref{eqn:forwardimage}-\eqref{eqn:odesolve}. If ground truth
$\mathbf{x}_{1:T}$ are available, 
we can optimize HyKE-Net with a 
supervised objective as the  mean squared error (MSE) between  $\mathbf{x}_{1:T}$ and $\hat{\mathbf{x}}_{1:T}$: 
\begin{equation}
\label{eqn:loss_s}
    \arg\min_{\theta,\phi,\varphi}\mathcal{L}_s = \left \| \mathbf{x}_{1:T} - \hat{\mathbf{x}}_{1:T} \right \|_2^2 + \lambda \left \| \hat{\mathbf{x}}_{1:T} - \tilde{\mathbf{x}}_{1:T} \right \|_2^2
\end{equation}
where the second regularization term 
provides additional constraints for $\mathbf{\tilde{x}}_{1:T}$ obtained in Equation \eqref{eqn:BPF} to be similar to reconstructed images $\hat{\mathbf{x}}_{1:T}$. 
If ground-truth 
$\mathbf{x}_{1:T}$ are not available,
we further  
%For unsupervised reconstruction, we need to 
obtain the reconstructed sinograms $\Bar{\mathbf{y}}_{1:T}$ from $\hat{\mathbf{x}}_{1:T}$ using the projection and measurement model as described in Equation \eqref{eq:projection}, %-\eqref{eq:measurement},
% \begin{equation}
%     \label{eqn:decoder} \Bar{\mathbf{y}}_{1:T}=\mathbf{D}(\mathbf{G}\hat{\mathbf{x}}_{1:T})+\mathbf{n}=\mathbf{D}\hat{\mathbf{y}}_{1:T}+\mathbf{n}
% \end{equation}
and formulate the unsupervised objective as the negative Poisson log-likelihood 
on $\Bar{\mathbf{y}}_{1:T}$  with the same regularization term used in Equation \eqref{eqn:loss_s}:
\begin{equation}
    \arg\min_{\theta,\phi,\varphi} \mathcal{L}_u = 
    -(\mathbf{y}_{1:T}^r \log \Bar{\mathbf{y}}_{1:T} - \Bar{\mathbf{y}}_{1:T}) + \lambda\left \| \hat{\mathbf{x}}_{1:T} - \tilde{\mathbf{x}}_{1:T} \right \|_2^2
\end{equation}

We choose the use of physics-based functions to bridge between the sinogram and image space, both in encoding (Equation \eqref{eqn:BPF}) and decoding (Equation \eqref{eq:projection}), to avoid the need of a large neural network for accommodating the size of PET images and sinograms (%\textcolor{blue}{
$128\times128$ and $160\times128$,  respectively, 
in our experiments).
%}).
% \hl{can yo give some rough sense of the size of these two things}

%In our experiments, we ablate the effect of allowing the encoding from sinograms $\mathbf{y}_{1:T}^r$ (Equation \eqref{eqn:BPF}) and decoding to sinograms $\Bar{\mathbf{y}}_{1:T}$ (Equation \eqref{eqn:decoder}) neural instead of physics-based.

% \hl{You should talk about 1) from kinetics to activity images (your current eq 14-15; and 2). from activity to sinogram (your current imaging part), which in the end summarize into your loss [can still give supervised alterantive as a bonus] }

\section{Experiments and Results}
\subsection{Experiments on Synthetic Data}
\subsubsection{Data:} 
We used the $128\times128$ 2D Zubal brain phantom \cite{phantom} with a tumor added. 
The phantom contained 6 regions of interests (ROIs) and the scanning schedule included 18 frames over 60 min as $3\times 60$ s, $9\times 180$ s, $6\times 300$ s. 
A two-tissue compartmental model with Feng’s input function \cite{Cp} was used to generate ground-truth kinetic $^{18}\text{F-FDG}$ scans, with its kinetic parameters selected randomly from Gaussian distributions as described in \cite{learning2}.
% selected randomly from Gaussian distributions \hl{what what mean and std}. 
The projection processes were performed following \cite{torchradon}, with 160 projection angles, 128 detector bins, 
20\% Poisson random noise, 
and a total photon-count of around $1.8 \times 10^7$. 
Scattered photons were not considered.
%was added to the projected sinograms, 
%Scattered photons were not considered,an. %the total photon-count used was around $1.8 \times 10^7$. 
We generated a total of 400 pairs of tracer activity images and sinograms: 320 for training, 40 for validation, and 40 for testing. 

\subsubsection{Models and Metrics:} 
We utilized one-tissue compartment model as described in Equation \eqref{eqn:1cmpt} as prior physics $f_{\text{PHY}}$ %. Note that the \textit{a priori} kinetics model 
to induce an inherent gap to the data-generating two-tissue compartment model. The neural kinetic function $f_{\text{NN}_\theta}$, spatial filter $h_\phi$, and kinetics encoder $g_\varphi(\cdot)$ were modeled as a simple fully-connected network, a residual CNN \cite{rCNN}, and a 3D-Unet \cite{3dUnet} respectively. 
For quantitative evaluation, we considered peak signal-to-noise ratio (PSNR) and  structural similarity (SSIM) for the activity image and MSE in the ROIs.
% \hl{of what} in all ROIs.
% \hl{you should prepare a supplemental material listing the architecture of the 3 components.}
% \hl{a bit more architecture details.} \hl{add descriptions of metrics used}

\subsubsection{Comparison with Existing Baselines:}
\begin{figure}[t]
    \centering
    \includegraphics[width=.9\linewidth]{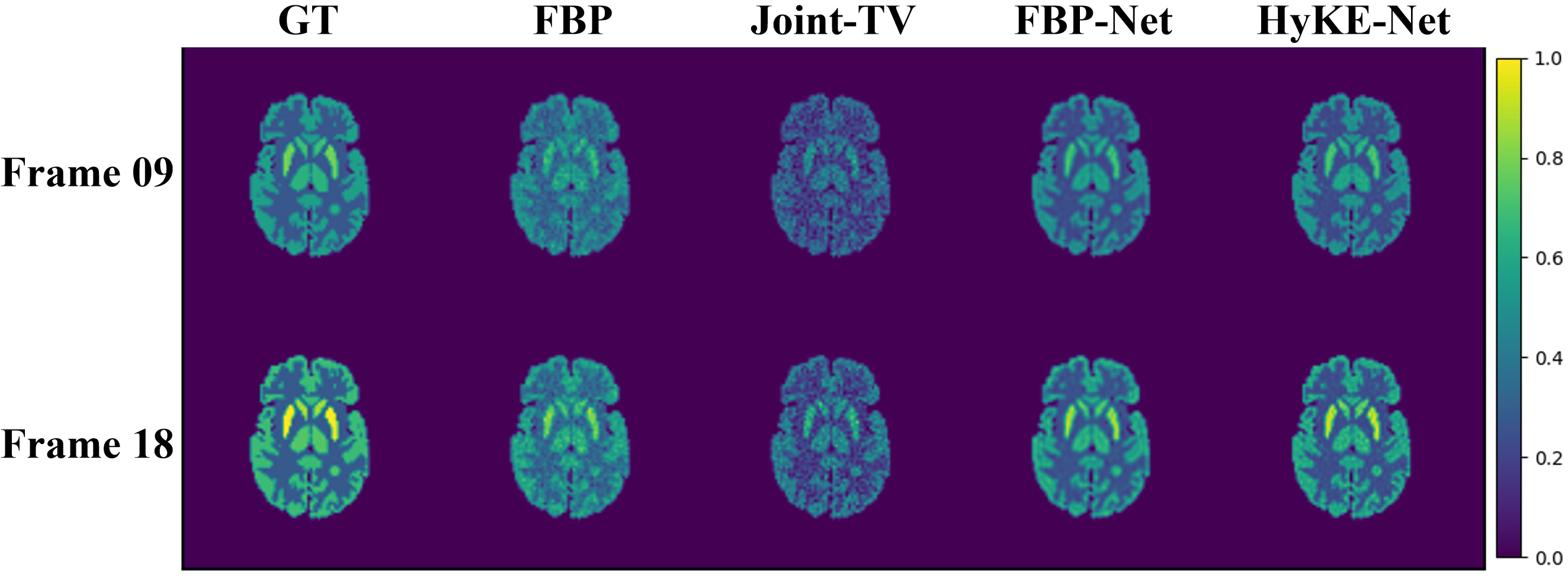}
    \caption{Examples of unsupervised reconstruction results from HyKE-Net and baselines.}
    \label{fig:ep1}
\end{figure}

We compared HyKE-Net with three existing methods including traditional direct reconstruction (FBP) \cite{FBP}, traditional kinetics-constrained reconstruction (Joint-TV) \cite{kinetic3}, and deep learning based reconstruction with spatial-temporal neural networks (FBP-Net) \cite{learning2}. While FBP-Net is supervised in its original formulations, we added our decoder in Equation \eqref{eq:projection} to test its performance in unsupervised reconstruction in comparison to the rest of the models. 
Quantitative results are summarized in Table \ref{tab1} and visual examples of all ROIs are provided in Fig.~\ref{fig:ep1}. While HyKE-Net's performance was moderately stronger than FBP-Net in supervised settings, it significantly outperformed all baselines in unsupervised settings. These suggest that, when ground-truth tracer activity images are not available to supervise the training, the use of hybrid kinetics brings significant benefits compared to purely physics-based (as in Joint-TV) or data-driven kinetics (as in FBP-Net). 

% \hl{add citation}
\begin{table}[t]
\caption{Comparison of HyKE-Net with baselines on phantom data.}
\label{tab1}
\centering
\resizebox{\linewidth}{!}{
\begin{tabular}{|c|ccc|ccc|}\hline
\multicolumn{1}{|l|}{\multirow{2}{*}{}} & \multicolumn{3}{c|}{Unsupervised Reconstruction}         & \multicolumn{3}{c|}{Supervised Reconstruction} \\ % \cline{2-7} 
\multicolumn{1}{|l|}{} & \multicolumn{1}{c}{MSE(e$^{-2}$) $\downarrow$} & \multicolumn{1}{c}{PSNR $\uparrow$} & \multicolumn{1}{c|}{SSIM $\uparrow$} & \multicolumn{1}{c}{MSE(e$^{-2}$) $\downarrow$} & \multicolumn{1}{c}{PSNR $\uparrow$} & \multicolumn{1}{c|}{SSIM $\uparrow$} \\ \hline
FBP  & 2.25(0.00) & 12.85(0.00) & 0.41(0.00) & / & / & / \\
Joint-TV & 5.61(0.00) & 18.69(0.00) & 0.69(0.00) & / & / & / \\
FBP-Net & 1.43(0.01) & 23.76(0.31) &  0.91(0.00)
& 0.05(0.01) & 40.02(0.70) & 0.98(0.00) \\
HyKE-Net & \textbf{0.75(0.00)} & \textbf{27.90(0.35)} & \textbf{0.94(0.00)} & \textbf{0.03(0.01)} & \textbf{42.33(0.73)} & \textbf{0.98(0.00)} \\ \hline
\end{tabular}}
\end{table}

\iffalse
We compared HyKE-Net with the two types of  dynamic PET reconstruction methods: 
1) unsupervised reconstruction methods including 
direct FBP reconstruction (FBP), 
and \hl{a physics-based deep learning method based on estimating the parameters of \textit{a priori} kinetics (Joint-TV)}; 
and 2) supervised reconstruction methods including \hl{a data-driven deep-learning method based on learning the tracer kinetics (FBP-Net), and the deep learning based iterative reconstruction method (STDP-Net)}. While the latter two data-driven approaches are supervised in their original formulations, we added the same measurement model (Equation \eqref{eq:measurement}) to allow us to test their performance in unsupervised reconstruction in comparison to the rest of the models. 
\hl{1. add citation to each; 2. need better descriptions to differentiate the last two models.} 
\fi

\subsubsection{Ablation of the Benefits of Hybrid Kinetics:}
\begin{figure}[t]
    \centering
    \includegraphics[width=.9\linewidth]{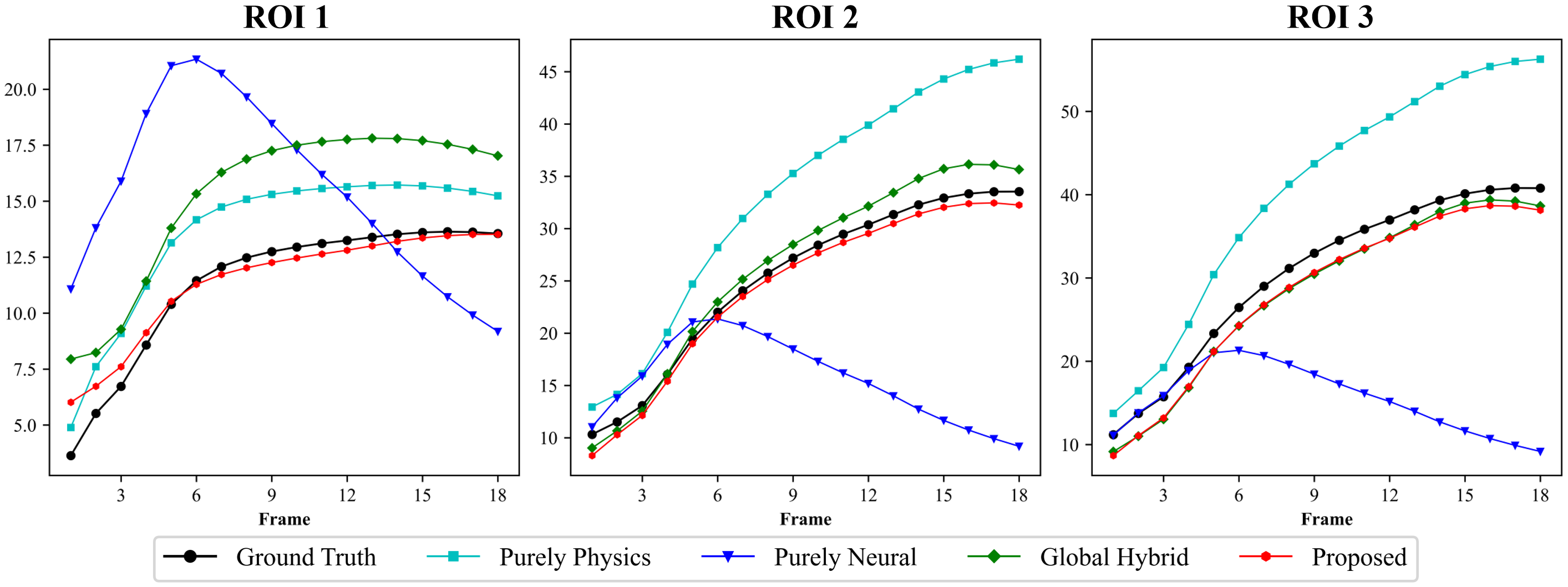}
    \caption{Examples of unsupervised reconstruction of time activity curves.}
    \label{fig:ep2}
\end{figure}

We further ablated the benefits of embedding hybrid kinetics by comparing the proposed HyKE model with the following three alternatives %kinetics model in the proposed HyKE-Net framework 
while keeping the rest of the framework identical: 1) purely physics kinetics: $\frac{d\mathbf{z}_{it}}{dt}=f_{\text{PHY}}(\mathbf{z}_{it}, a_t; \mathbf{c}_{pi})$; 2) purely neural kinetics: $\frac{d\mathbf{z}_{it}}{dt}=f_{\text{NN}_\theta}(\mathbf{z}_{it};\mathbf{c}_{ni})$; and 3) global hybrid kinetics: $\frac{d\mathbf{z}_{it}}{dt}=f_{\text{PHY}}(z_{it}, a_t; \mathbf{c}_{pi})+f_{\text{NN}_\theta}(\mathbf{z}_{it})$ where the neural component $f_{\text{NN}_\theta}$ is shared across all samples.  
%4) adaptive hybrid kinetics: $\frac{dz_{it}}{dt}=f_{\text{PHY}}(z_{it}, a_t; c_{pi})+f_{\text{NN}_\theta}(z_{it};c_{ni})$ on simulation data. 
% Note that all physics-based kinetics model assumes the one-tissue compartment model, thus inherent with a gap to the data-generating two-tissue compartment model. 
Quantitative results are summarized in Table \ref{tab2} and examples of  unsupervised reconstruction of the time activity curve (TAC) in three ROIs are shown in Fig.~\ref{fig:ep2}. As demonstrated, both purely physics-based and purely neural kinetics struggle to reconstruct the data-generating kinetics. The use of global hybrid kinetics helps (green), but can be further improved by a hybrid kinetics model adaptable to the data as in HyKE to consider the heterogeneous errors in the prior physics.
%significantly improves over the alternative  p, or hybrid models with a 
% \subsubsection{Ablation of the Benefits of Encoder and Decoder}: In this experiment, we replace the encoder and decoder with purely neural network.

%Notable, there's a gap between the kinetics of data and prior kinetic model.
\begin{table}[t]
\caption{Ablating the benefits of hybrid kinetics with an adaptive neural component.}
\label{tab2}
\centering
\resizebox{\linewidth}{!}{
\begin{tabular}{|l|ccc|ccc|}\hline
\multicolumn{1}{|l|}{\multirow{2}{*}{}} & \multicolumn{3}{c|}{Unsupervised Reconstruction}         & \multicolumn{3}{c|}{Supervised Reconstruction} \\ % \cline{2-7} 
\multicolumn{1}{|l|}{} & \multicolumn{1}{c}{MSE(e$^{-2}$) $\downarrow$} & \multicolumn{1}{c}{PSNR $\uparrow$} & \multicolumn{1}{c|}{SSIM $\uparrow$} & \multicolumn{1}{c}{MSE(e$^{-2}$) $\downarrow$} & \multicolumn{1}{c}{PSNR $\uparrow$} & \multicolumn{1}{c|}{SSIM $\uparrow$} \\ \hline
Purely Physics & 5.97(0.36) & 18.89(0.27) & 0.63(0.03)
& 0.07(0.00) & 38.10(0.01) & 0.97(0.00) \\
Purely Neural & 19.55(0.21) & 13.74(0.05) &  0.61(0.01)
& 0.04(0.01) & 41.14(0.12) & 0.98(0.00) \\
Global Hybrid  & 1.39(0.04) & 25.22(012) &  0.92(0.02)
& 0.04(0.00) & 40.72(0.58) & 0.98(0.00) \\
HyKE-Net & \textbf{0.75(0.00)} & \textbf{27.90(0.35)} & \textbf{0.94(0.00)} & \textbf{0.03(0.01)} & \textbf{42.33(0.73)} & \textbf{0.98(0.00)} \\ \hline
\end{tabular}}
\end{table}

\subsection{Experiments on Real Data}

\subsubsection{Data \& Models:} 
Animal data were obtained from those described in \cite{learning2}. 
% \hl{cite the paper where the data comes from}, 
%experiments were approved by the experimental animal welfare and ethics review committee, and were performed in compliance with local legal requirements. 
%where 10 rats with gliomas were anesthetized and injected with 1 mCi FDG before 60-minute dynamic PET scans by Siemens Micro-PET/CT Inveon. 
The sampling protocol for dynamic PET scans was set to 3 × 60 s, 9 × 180 s, 6 × 300 s. 
The ground-truth $\mathbf{x}_{1:T}$ were reconstructed with CT attenuation correction and full 3D counts.
Due to the lack of access to the CT images and thus the inability to properly model the actual sinogram measurement process, 
we 
simulated low-count sinograms with $2.4 \times 10^5$ photon-count with different levels of randoms and scatters: we tested how HyKE-Net and its baselines responds to these noises in the data when the measurement model is unaware of these noies. 
Note that the data-generating tracer kinetics underlying $\mathbf{x}_{1:T}$ is not known, which is the main motivation for HyKE-Net.
%and the attenuation was not considered due to the lack of access to real CT images. 
% For ground-truth $\mathbf{x}_{1:T}$ needed for supervised-learning, reconstructions with CT attenuation correction and full 3D counts. 
% Unfortunately, we did not have access to the CT images in this dataset and thus lacked the ability to properly model the sinogram measurement process. 
% Instead, we simulated low-count sinograms with $2.4 \times 10^5$ total photon-count, \hl{and introduced various errors in the measurement model} (Equations \eqref{eq:projection}-\eqref{eq:measurement}) used in HyKE-Net to mimic real-data use: 
% \hl{describe what errors being tested. 
% Because the real data-generating kinetics is unknown, 
%we further test the use of one-tissue \textit{vs}. two-tissue compartment models in HyKE-Net.}
600 images from 10 rates were used, with eight rats selected randomly for training, one for validation, and one for testing.

%The 3D sinograms (michelograms, 128 bins, 160 views) of each rat can be reconstructed into 18 frames of 3D images, each frame contains 159 slices of size 128×128 and the reconstructed PET images with full 3D counts and CT attenuation correction were used as labels. 
% and only the segments 0 of the michelograms were used as training and testing 2D sinograms \hl{I did not understand this part}. The reconstructed PET images with full 3D counts and CT attenuation correction were used as labels \hl{for the scenario of supervised training.} 
%The experiment contains data from a total of 10 rats.\hl{why 12 earier?}

%\subsubsection{Models:} 
Since real $C_p(t)$ (or $a_t$) is not available in real data,
we used $C_p$ from synthetic-data experiments but also provided it as an input to the neural kinetic function %$f_{\text{NN}_\theta}(\mathbf{z}_{it};\mathbf{c}_{ni})$ 
$f_{\text{NN}_\theta}(\mathbf{z}_{it}, a_t;\mathbf{c}_{ni})$ to allow the modeling of its errors. 
%and  \hl{does the compartment model still takes cp as input?} \textcolor{blue}{Yes, but the simulated Cp}. 
We compared HyKE-Net with existing baselines for unsupervised reconstruction considering 
10\% \textit{vs}.\ 20\% randoms and scatters 
in the measured sinograms. % 10\% randoms \& 10\% scatters, and 20\% randoms \& 20\% scatters. 
Because the data-generating kinetics is unknown, 
we further tested the effect of using the one-tissue \textit{vs}. two-tissue compartment model as prior physics in HyKE-Net (noted as HyKE-Net1 and HyKE-Net2 respectively).
% \subsubsection{Effect of Prior Physics in Hybrid Kinetics:}

\begin{table}[t]
\caption{Comparison of HyKE-Net with baselines on real data.}
\label{tab3}
\centering
\resizebox{\linewidth}{!}{
\begin{tabular}{|c|ccc|ccc|}\hline
\multicolumn{1}{|l|}{\multirow{2}{*}{}} & \multicolumn{3}{c|}{10\% randoms and 10\% scatters}     & \multicolumn{3}{c|}{20\% randoms and 20\% scatters} \\ % \cline{2-7} 
\multicolumn{1}{|l|}{} & \multicolumn{1}{c}{MSE(e$^{-3}$) $\downarrow$} & \multicolumn{1}{c}{PSNR $\uparrow$} & \multicolumn{1}{c|}{SSIM $\uparrow$} & \multicolumn{1}{c}{MSE(e$^{-3}$) $\downarrow$} & \multicolumn{1}{c}{PSNR $\uparrow$} & \multicolumn{1}{c|}{SSIM $\uparrow$} \\ \hline
FBP  & 80.00(0.00) & 10.97(0.00) & 0.03(0.00) & 84.05(0.00) & 10.75(0.00) & 0.03(0.00) \\
Joint-TV & 1.73(0.00) & 27.63(0.00) & 0.31(0.00) & 1.84(0.00) & 27.35(0.00) & 0.31(0.00) \\
FBP-Net & 0.24(0.02) & 36.24(0.42) & 0.60(0.01) 
& 0.42(0.05) & 33.82(0.48) & 0.55(0.03) \\
HyKE-Net1 & 0.23(0.01) & 36.12(0.68) & 0.61(0.00) & \textbf{0.26(0.02)} & \textbf{36.93(0.43)} & \textbf{0.60(0.01)} \\ 
HyKE-Net2 & \textbf{0.22(0.02)} & \textbf{36.71(0.50)} & \textbf{0.61(0.00)} & 0.32(0.05) & 34.91(0.43) & 0.58(0.00) \\
\hline
\end{tabular}}
\end{table}

\begin{figure}[t]
    \centering
    \includegraphics[width=.9\linewidth]{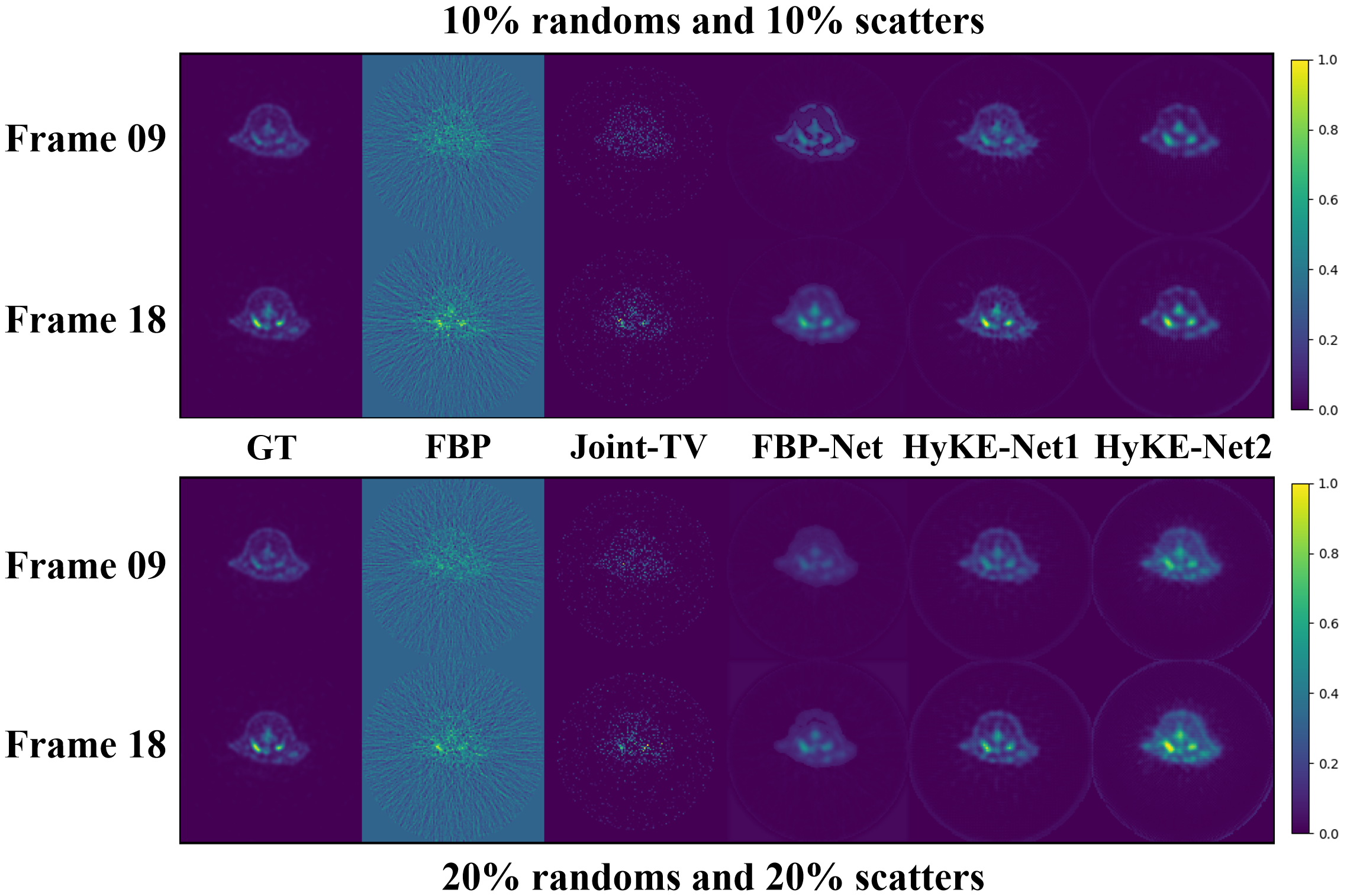}
    \caption{Examples of unsupervised reconstruction results on real data.}
    \label{fig:ep3}
\end{figure}

\subsubsection{Results:} The quantitative results are summarized in Table \ref{tab3} with visual examples shown in Fig.~\ref{fig:ep3}. 
While all methods' performance decreased as the noises in the sinograms (and thus the errors in the measurement model) increased, HyKE-Net consistently outperformed its baselines in both settings. 
In particular, 
when using the one-tissue compartment model as the prior physics, 
HyKE-Net demonstrated a strong robustness to measurement errors compared to all baselines. 
Interestingly, when using a more sophisticated two-tissue compartment model, 
the performance of HyKE-Net showed marginal improvements given lower measurement errors but faster deterioration given higher measurement errors -- this may suggest that, when measurement quality is low, a more complex physics-based kinetics model may increase the difficulty of identifiability. 

\section{Conclusion}
We present a novel framework HyKE-Net for dynamic PET reconstruction that embeds hybrid tracer kinetics into a physics-based encoding-decoding architecture. Experiment on synthetic data and real data demonstrated its benefits from integrating rich prior knowledge while allowing for its errors, especially in unsupervised dynamic PET reconstruction. Future work will focusing on addressing the increased computational cost of HyKE-Net due to hybrid modeling, 
and the impact on unsupervised reconstruction by the errors in the measurement model -- including the possibility to make the measurement model hybrid. 
%has a greater computational cost compared to other deep learning methods and its performance may be  and 
Future works will also focus on extensive quantitative evaluations on real human data.

\hspace*{\fill} \

\noindent\textbf{Acknowledgements}. This work is supported in part by the National Key Research and Development Program of China(No: 2020AAA0109502); the Talent Program of Zhejiang Province (No: 2021R51004); NIH NHLBI grant R01HL145590 and NSF OAC-2212548.

%
% ---- Bibliography ----
%
% BibTeX users should specify bibliography style 'splncs04'.
% References will then be sorted and formatted in the correct style.
%
% \bibliographystyle{splncs04}
% \bibliography{mybibliography}
%
\bibliographystyle{splncs04}
\bibliography{reference}
\end{document}